\documentclass[aps,prb,twocolumn,superscriptaddress,showpacs]{revtex4-1}

\usepackage{graphicx}
\usepackage{calc}
\usepackage{bm}
\usepackage{color}
\usepackage{mathtools}

\begin{document}

\title{Altermagnetic  bulk and topological surface magnetizations for CrSb single crystals}

\author{N.N.~Orlova}
\author{A.A.~Avakyants}
\author{V.D.~Esin}
\author{A.V.~Timonina}
\author{N.N.~Kolesnikov}
\author{E.V.~Deviatov}
\affiliation{Institute of Solid State Physics of the Russian Academy of Sciences, Chernogolovka, Moscow District, 2 Academician Ossipyan str., 142432 Russia}

\date{\today}

\begin{abstract}
We experimentally investigate the angle dependence of magnetization   $M(\alpha)$ for single crystals of CrSb. CrSb belongs to a new class of altermagnetic materials, the small net magnetization is accompanied by alternating spin splitting in the k-space. In addition,  CrSb reveals also topological features with  Weyl surface states originating from bulk band topology. We observe, that $M(\alpha)$ oscillates around zero value, so magnetization is positive for $M(\alpha)$ maxima and it is negative for $M(\alpha)$ minima. The magnetization reversal curves $M(H)$ are non-linear with low-field hysteresis, but with almost linear high-field branches. The slope of the linear branches well correlates with $M(\alpha)$ oscillations, so it is positive for $M(\alpha)$ maxima and negative for $M(\alpha)$ minima. We demonstrate, that the interplay between the positive and the negative $M(H)$ slopes originates from several magnetic phases in CrSb. In particular, current-carrying topological surface states are responsible for the diamagnetic-like $M(H)$ negative slope, which dominates for the directions of full spin compensation in the bulk CrSb altermagnetic spectrum. Due to the spin-momentum locking, topological surface states are spin-polarized, which is responsible for the  low-field hysteresis.  Thus, we experimentally  demonstrate both the altermagnetic  bulk and the topological surface magnetizations for the  altermagnetic candidate CrSb. 
  
\end{abstract}

\maketitle

\section{Introduction}

Recently, the concept of spin-momentum locking~\cite{Armitage,sm-valley-locking} was extended to the case of weak spin-orbit coupling, i.e. to the non-relativistic groups of magnetic symmetry~\cite{alter_common1, alter_common2, alter_mazin}. As a result, the small net magnetization is accompanied by alternating spin-splitting, forming d-, g- or i-wave altermagnetic symmetry: the oppositely  spin-polarized subbands are connected by rotation  in the k-space, in contrast to conventional anti- and ferromagnets~\cite{alter1,alter2}. For example, for the d-wave order parameter, the up-polarized subband can be obtained by $\pi/2$ rotation of the down-polarized one  in the k-space~\cite{alter_supercond_notes,alter_normal_junction}.

Despite numerous theoretical predictions, only a few altermagnetic candidates are verified in experiment. For example, ARPES (angular-resolved photoemission spectroscopy) and SX-ARPES (spin-integrated soft X-ray angular-resolved photoemission spectroscopy) confirm the altermagnet nature of spin-spliting in $\alpha$-MnTe~\cite{alter1, MnTe_ARPES2}. Anomalous Nerst (ANE) and Hall (AHE) effects~\cite{Armitage,alter_original} have been  been experimentally demonstrated~\cite{AHE_RuO2,AHE_MnTe1,AHE_MnTe2,AHE_Mn5Si3,Mn5Si3_1,Mn5Si3_3} for most of altermagnetic candidates.  

It is a common  agreement, that AHE still requires  spin-orbit coupling even in altermagnetic  materials~\cite{AHE_MnTe1,AHE_MnTe2,spin_ferro_soc,orbital_mag1}. For MnTe altermagnet, the principle origin of finite net  magnetization~\cite{alter_ferro,spin_ferro_soc,orlova_MnTe1,orlova_MnTe2} is the spin-orbit coupling~\cite{satoru} in valence orbitals~\cite{Dichroism}.  The effects of spin-orbit coupling in this material have been demonstrated by temperature-dependent ARPES~\cite{MnTe_SO} and by magnetization measurements~\cite{orlova_MnTe1}.

In contrast to MnTe, spin-orbit coupling is weak and plays a minor role in the low energy band structure for CrSb altermagnetic candidate~\cite{ARPES2_CrSb}, while ARPES confirms high altermagnetic splitting up to 1.0~eV near the Fermi energy~\cite{ARPES1_CrSb, ARPES2_CrSb}.  CrSb is of the same crystal point group symmetry as  $\alpha$-MnTe, i.e.  CrSb is of hexagonal structure with the space group $P6_3/mmc$  (No. 194) and with the magnetic space group  $P6'_3/m'm'c$. In CrSb,  two Cr sublattices with opposite spins are aligned along the $c$ axis, being connected by $C6z$  -  a six-fold rotation  combined with  a 1/2 translation along c~\cite{neudiff1, neudiff2}. The N\'eel temperature is about 705~K, which might be convenient in applications. Thus, it is reasonable to investigate the CrSb  magnetization in a wide field range, to compare with the spin-orbit-induced effects in MnTe altermagnet~\cite{orlova_MnTe1, orlova_MnTe2}.  

The concept of spin-momentum locking~\cite{sm-valley-locking} was originally proposed for magnetic and non-magnetic topological materials, which are characterized by the topological surface states originating from bulk band topology~\cite{Volkov-Pankratov,MZHasan,Armitage}. Recently, chiral Weyl surface states have also been theoretically proposed for altermagnets~\cite{AMtopology1,AMtopology2}.  The altermagnetic  band spin splitting and signature of topological surface states were shown for (100) cleaved CrSb surface by ARPES~\cite{Weyl alter1_CrSb}. Thus, room-temperature altermagnetic candidate CrSb reveals both altermagnetic and topological features~\cite{Weyl alter2_CrSb}, which might also affect the single-crystal sample magnetization~\cite{cosns}.

Here, we experimentally investigate magnetization behavior for single crystals of altermagnetic candidate CrSb. The $M(\alpha)$ angle dependence of magnetization    oscillates around zero value, so magnetization is positive for $M(\alpha)$ maxima and it is negative for $M(\alpha)$ minima. The magnetization reversal curves $M(H)$ are non-linear in low magnetic fields with unusual hysteresis  within $\pm 3$~kOe, but with linear high-field branches with positive ($M(\alpha)$ maxima) or negative ($M(\alpha)$ minima) slopes.  We also do not observe $\pi/3$ periodicity of magnetization for any sample orientation, despite it could be expected for the hexagonal CrSb structure.

\begin{figure}
\includegraphics[width=1\columnwidth]{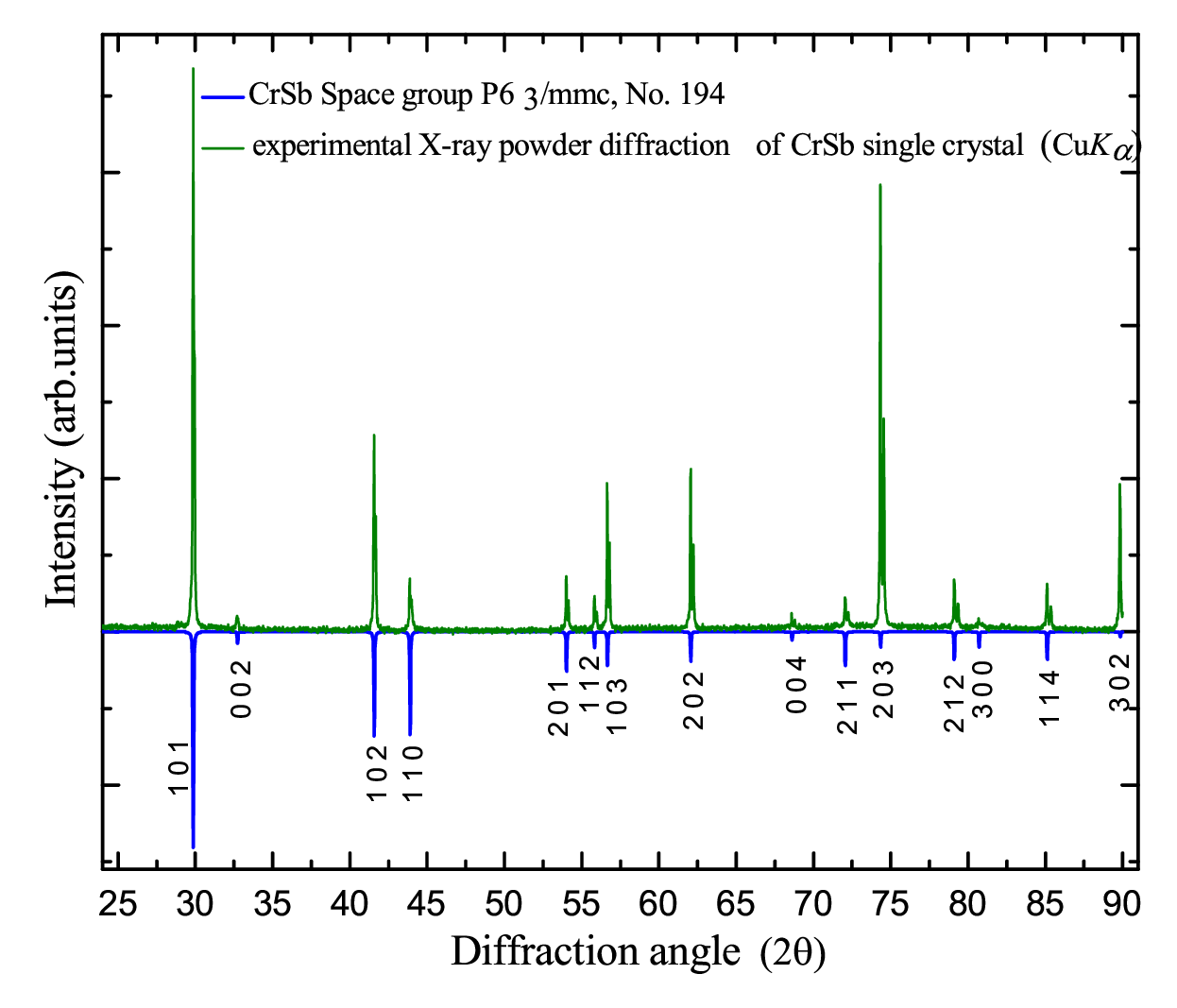}
\caption{(Color online) (a)  The X-ray powder diffraction  pattern (Cu K$_{\alpha1}$ radiation), which is obtained for the crushed CrSb single crystal. The single-phase  CrSb is confirmed with the space group $P6_3 /mmc$ No. 194. 
  }
\label{sample}
\end{figure}

\section{Samples and technique}

CrSb single crystals were synthesized by reaction of elements. Cr (99.996\%) and Sb (99.9999\%) were mixed in the stoichiometric ratio and then heated in an evacuated silica ampule up to 1000$^\circ$C with the rate of 15$^\circ$C/h in a gradient-free furnace. The load was held at 1000$^\circ$C for 72 hours and then cooled down slowly (11$^\circ$C/h) to the room temperature. The crystals grown are faceted single crystals with the space group $P6_3 /mmc$ No. 194 and the stoichiometric composition, as confirmed by X-ray diffraction analysis, see Fig.~\ref{sample}.

To investigate angle-dependent magnetization, it is preferable to use small CrSb single crystal samples rather than thin films. In the latter case, the results may be seriously affected by the geometrical factors and by the admixture of the substrate magnetic response, especially in low magnetic fields. Also, the presence of the original crystal facets might be important in view of possible surface effects~\cite{Weyl alter1_CrSb,Weyl alter1_CrSb}. For these reasons, we choose for investigation small CrSb single crystal samples with different masses (1.39~mg and 2.75~mg, respectively), which allows to variate the surface to bulk ratio. The samples are taken 'as grown', without any treatment of the sample surface. 

\begin{figure}
\includegraphics[width=\columnwidth]{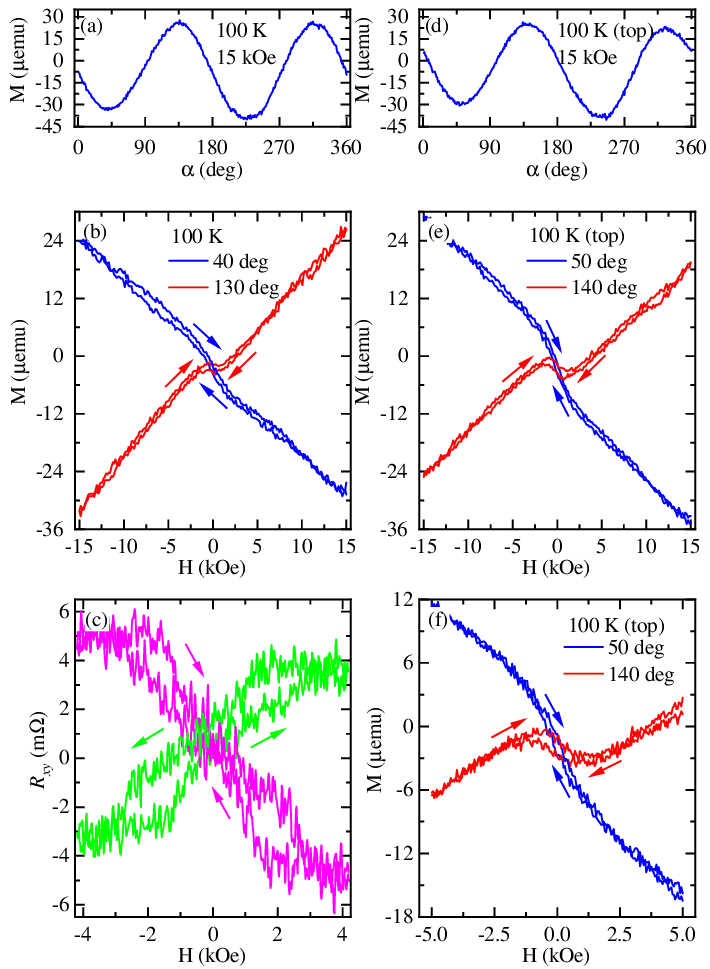}
\caption{(Color online)  (a) $M(\alpha)$ angle dependence of magnetization for the smallest, 1.39~mg, CrSb single crystal sample. The curve is obtained for 15~kOe magnetic field at 100~K temperature, the sample is mounted to the side sample holder plane. 
(b) $M(H)$ magnetization loops  for $\alpha$ = 40$^\circ$ ($M(\alpha)$ minimum) and for $\alpha$ = 130$^\circ$ ($M(\alpha)$ maximum) the blue and the red curves, respectively. For both angles, $M(H)$ is non-linear with low-field ($\pm 3$~kOe) hysteresis, but with linear diamagnetic-like (the blue curves) or paramagnetic-like (the red ones) high-field branches.
(c) Anomalous Hall effect  measurements for small single-crystal CrSb flake at 1.2~K. The $R_{xy}(H)$ curves show low-field $\pm 3$~kOe hysteresis, the  $R_{xy}(H)$ slope changes a sign for two $20^\circ$ rotated Hall-bar configurations for the same flake. 
(d-f) Qualitatively similar to (a-b) curves for the top mount position of the sample.
(f) The region of the low-field ($\pm 3$~kOe)  $M(H)$ hysteresis is always characterized by negative (diamagnetic-like) $M(H)$ slope, even for the the $M(\alpha)$ maxima (the red curves). 
}
\label{fig2}
\end{figure} 

Sample magnetization  is measured by Lake Shore Cryotronics 8604 VSM magnetometer, equipped with nitrogen flow cryostat.  Angle-dependent magnetization $M(\alpha)$ is investigated by sample holder rotation in  magnetic field.   Also, the sample can be mounted to either the side or the top sample holder planes, to variate CrSb crystal orientation in respect to the magnetic field and the rotation axis~\cite{orlova_MnTe2}. The sample is mounted to the sample holder by low temperature grease. It was verified~\cite{gete}, that without a sample, the sample holder with corresponding amount of grease shows fully isotropic and strictly linear small diamagnetic response, which can be estimated as below  10\% of the measured CrSb magnetization value. Thus, the experimental setup allows to obtain magnetization angle dependence with high resolution in a wide magnetic field range $\pm15$~kOe. 

Before any measurements, the sample is cooled down the minimal 78~K temperature in zero magnetic field.  Afterward, the sample  is magnetized at 15~kOe, to have the stable, well-reproducible initial sample state.

\section{Experimental results}

$M(\alpha)$ angle dependence of magnetization is demonstrated in Fig.~\ref{fig2} (a) for the smallest, 1.39~mg, CrSb single crystal sample. The $M(\alpha)$ curve is obtained for 15~kOe magnetic field at 100~K temperature, the sample is mounted to the side sample holder plane.  Two maxima ($\alpha$ = 130$^\circ$ and $\pi$-shifted) and two minima ($\alpha$ = 40$^\circ$ and $\pi$-shifted) indicate  $\pi$ periodicity of the experimental $M(\alpha)$ curve, which does not correspond to the $\pi/3$ hexagonal CrSb crystalline symmetry.

As the main experimental result, $M(\alpha)$ oscillates around zero value in Fig.~\ref{fig2} (a), so magnetization is positive for $M(\alpha)$ maxima while it is definitely negative for $M(\alpha)$ minima. In our setup,  $M(\alpha)$ is not allowed to change a sign for homogeneous, single-phase sample magnetization: the magnetometer detector coils are fixed to the magnet pole caps (not to the sample holder), so, in finite external magnetic field, magnetization is defined by the field direction: it is always positive for the ferromagnetic~\cite{cosns} or antiferromagnetic~\cite{orlova_MnTe1, orlova_MnTe2} samples, or negative for the diamagnetic ones~\cite{gete}. Only the remanence magnetization is allowed to change a sign (the rotating sample does not change magnetization in zero external field), while Fig.~\ref{fig2} (a) shows positive and negative magnetization values in 15~kOe magnetic field.  Thus, oscillations in Fig.~\ref{fig2} (a) require several magnetic phases in CrSb single crystal sample. 

This conclusion is illustrated by direct measurements of $M(H)$ magnetization reversal curves in Fig.~\ref{fig2} (b) for $\alpha$ = 40$^\circ$ ($M(\alpha)$ minimum) and for $\alpha$ = 130$^\circ$ ($M(\alpha)$ maximum), the blue and the red curves, respectively. For both angles, $M(H)$ curves are non-linear in low magnetic fields ($\pm 3$~kOe), but with almost linear diamagnetic-like (the blue curves) or paramagnetic-like (the red ones)  high-field branches. Usually, small high-field $M(H)$ hysteresis is due to the antiferromagnetic domain configuration change, the low-field one could also be expected for altermagnets~\cite{orlova_MnTe1, orlova_MnTe2,weak_ferro,weak_ferro1}. However,  the interplay between the para- and diamagnetic high-field behavior for different angles is impossible  for materials with single magnetic phase. 
 
The observed interplay can be confirmed by anomalous Hall effect (AHE) measurements for the small single-crystal CrSb flake, see Fig.~\ref{fig2} (c) (the details of the sample preparation and the measurement technique will be published elsewhere). The AHE transverse current is usually assumed to be perpendicular to  magnetization, therefore,  AHE and sample magnetization share the same symmetry~\cite{satoru}. The $R_{xy}(H)$ curves show low-field $\pm 3$~kOe hysteresis, the  $R_{xy}(H)$ slope changes a sign in Fig.~\ref{fig2} (c) for two $20^\circ$ rotated Hall-bar configurations at the same flake. Since the sign of the charge carriers is the same, we independently confirm the interplay between  positive and negative $M(H)$ slopes in Fig.~\ref{fig2} (a) and (b). 

These results can be reproduced for the top mount position of the sample, see  Fig.~\ref{fig2} (d-f). The rotation axis is therefore turned 90$^\circ$ in respect to the previous case, but the behavior is qualitatively the same: $M(\alpha)$ oscillates around zero value in Fig.~\ref{fig2} (d), so $M(H)$ is changed from the diamagnetic-like at $M(\alpha)$  minima (e.g. for $\alpha$ = 50$^\circ$, the blue curves) to the paramagnetic-like at maxima ($\alpha$ = 140$^\circ$, the red curves), see Fig.~\ref{fig2} (e). There is low-field  $M(H)$ hysteresis within $\pm 3$~kOe as depicted in Fig.~\ref{fig2} (f). The hysteresis  region is always characterized by negative (diamagnetic-like) $M(H)$ slope, even for the the $M(\alpha)$ maxima (the red curves), which is quite unusual for standard magnetic ordering~\cite{orlova_MnTe1, orlova_MnTe2,weak_ferro,weak_ferro1}. Despite the shape of the hysteresis is different for AHE in Fig.~\ref{fig2} (c) and for $M(H)$ in Fig.~\ref{fig2} (f), it is within the same $\pm 3$~kOe magnetic field range.

\begin{figure}
\includegraphics[width=\columnwidth]{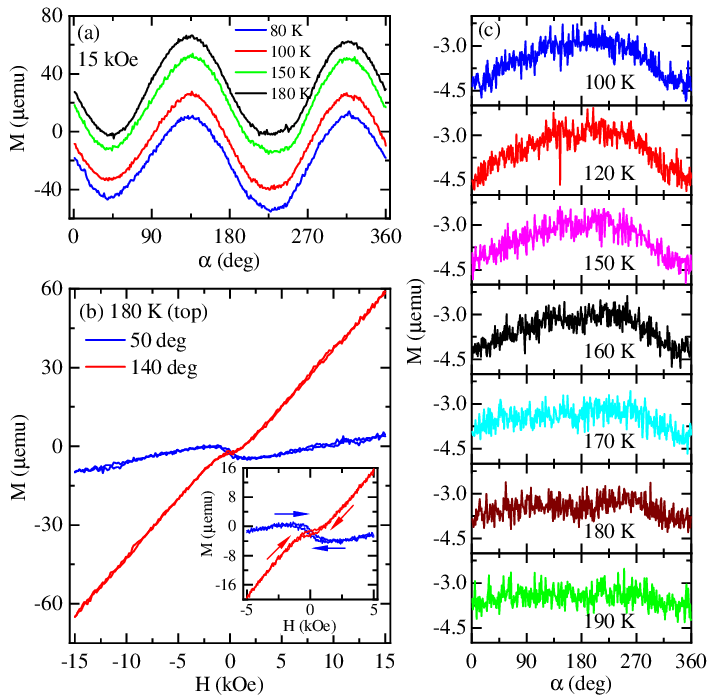}
\caption{(Color online) 
(a) $M(\alpha)$ temperature dependence for the smallest, 1.39~mg, CrSb single crystal sample. The curves are obtained for the top mount position of the sample for 15~kOe magnetic field.  The oscillations' amplitude is nearly independent of  temperature, while the whole  $M(\alpha)$ curve is rising for higher temperatures.
(b) $M(H)$ magnetization loops at 180~K  for $\alpha$ = 50$^\circ$ ($M(\alpha)$ minimum) and for $\alpha$ = 140$^\circ$ ($M(\alpha)$ maximum), the blue and the red curves, respectively. The slopes of the high-field branches are simultaneously increased, conserving vertical distance between the branches. Inset shows the low-field hysteresis, which is not sensitive to temperature. 
(c) Remanence $M(\alpha, H=0)$ magnetization.  At low temperatures, $M(\alpha, H=0)$ curves are $2\pi$ periodic, which well correspond to the easy axis magnetization, while the amplitude is diminishing with the temperature increase. 
 }
\label{fig3}
\end{figure}

Fig.~\ref{fig3} (a) shows temperature dependence of the observed $M(\alpha)$ oscillations. The oscillations' amplitude is nearly independent of  temperature, while the whole  $M(\alpha)$ curve is rising with increasing the temperature. This temperature dependence is reflected in $M(H)$ magnetization loops in Fig.~\ref{fig3} (b): while the low-field hysteresis is not sensitive to temperature (see also the inset), the slopes of the high-field branches are increased at 180~K for both $\alpha$ = 50$^\circ$ ($M(\alpha)$ minimum) and $\alpha$ = 140$^\circ$ ($M(\alpha)$ maximum), conserving vertical distance between the branches.  

The low-field hysteresis implies finite remanence $M(H=0)$ magnetization, which is shown in Fig.~\ref{fig3} (c). The curves are obtained in zero external magnetic field after magnetization of the sample at 15~kOe for the intermediate angle  $\alpha$ = 0$^\circ$. At low temperatures, $M(\alpha, H=0)$ curves are $2\pi$ periodic, which well corresponds to the easy axis magnetization: the rotating sample does not change magnetization in zero external field, which results in $2\pi$-periodicity. In contrast, in high magnetic fields,  magnetization is always aligned along the external field, demonstrating  $\pi$-periodicity for the same easy axis magnetization, see Fig.~\ref{fig3} (a). The $M(\alpha, H=0)$ amplitude is diminishing with temperature in Fig.~\ref{fig3} (c). Thus, low-field hysteresis can be attributed to the ferromagnetic-like spin ordering in the sample, despite negative $M(H)$ slope within $\pm 3$~kOe. 

\begin{figure}
\includegraphics[width=\columnwidth]{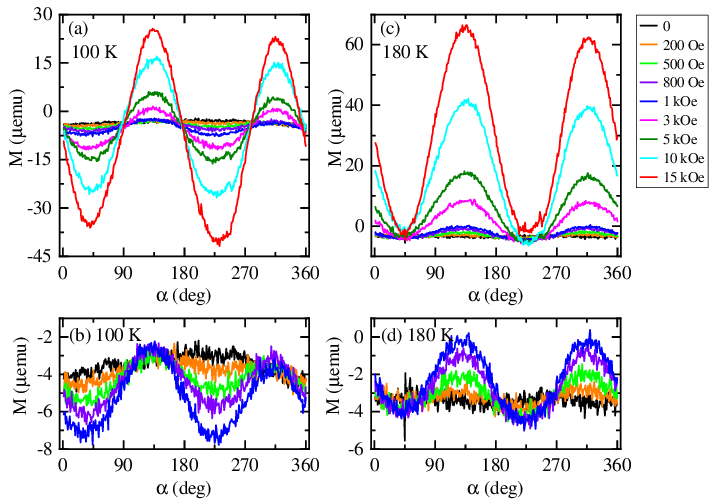}
\caption{(Color online) 
Angle dependence of magnetization $M(\alpha)$ for several magnetic fields  at 100~K (a,b) and at 180~K (c,d) temperatures.
The curves are obtained at 0, 200 Oe, 500 Oe, 800 Oe, 1~kOe, 3~kOe, 5~kOe, 10~kOe, 15~kOe fields for the top mount position of the 1.39~mg CrSb sample.  $M(\alpha)$ minima and maxima dependences are in good correspondence with the curves in  Figs.~\ref{fig2} (e) and.~\ref{fig3} (b). The enlarged regions (b) and (d)  show   $M(\alpha)$ for 0 -- 1~kOe fields.  For $M(\alpha)$ minima, the magnetization is still negative in low fields even at 180~K. 
 }
\label{fig4}
\end{figure}

Fig.~\ref{fig4} illustrates the above described results as the  angle dependence of magnetization $M(\alpha)$ for several magnetic fields at 100~K  (a,b) and at 180~K (c,d) temperatures.  At 100~K, $M(\alpha)$ minima and maxima simultaneously go to zero, in a good correspondence with  Fig.~\ref{fig2} (e). In contrast, at 180~K temperature, $M(H)$ is weakly dependent on   magnetic field for $\alpha$ = 50$^\circ$ ($M(\alpha)$ minimum) in Fig.~\ref{fig3} (b) (the blue curve), so magnetization is weakly sensitive to magnetic field in $M(\alpha)$ minima, see  Fig.~\ref{fig4} (c). We wish to emphasize, that  sample magnetization is  still negative in low fields even at 180~K temperature,  as depicted in Fig.~\ref{fig4} (d).

\begin{figure}
\includegraphics[width=\columnwidth]{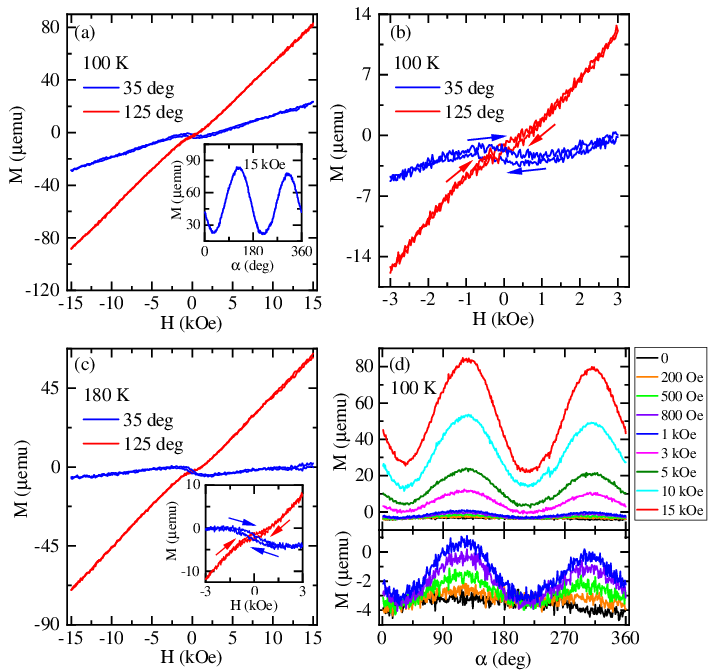}
\caption{(Color online) Similar magnetization results for  the larger $m = 2.75$~mg sample, the details depend on the sample mass.  
(a) $M(H)$ magnetization is non-linear in low fields, while the high-field linear branches are of different slopes for M($\alpha$) minima and maxima, e.g. for  $\alpha$ = 35$^\circ$ and  $\alpha$ = 125$^\circ$, respectively.  Inset confirms $\pi$-periodic $M(\alpha)$ angle dependence of magnetization. The curves are obtained at 100~K temperature.
(b) The low-field $M(H)$ region with  hysteresis within $\pm 2$~kOe  at 100~K.
(c) $M(H)$ magnetization at 180~K, inset shows the low-field  region. The slopes of the linear high-field branches are diminished in comparison with (a), in contrast to the smallest 1.39~mg  sample in Fig.~\ref{fig3}, so for $M(\alpha)$ minima, magnetization is always negative.
(d) Angle dependence of magnetization $M(\alpha)$ at 100~K for 0, 200 Oe, 500 Oe, 800 Oe, 1~kOe, 3~kOe, 5~kOe, 10~kOe, 15~kOe magnetic fields. The enlarged region shows $M(\alpha)$ for 0 -- 1~kOe fields.   For $M(\alpha)$ minima, magnetization is negative in low fields.
 }
\label{fig5}
\end{figure}

These results can be reproduced for different samples, however, the details depend on the sample mass. For example, Fig.~\ref{fig5} shows magnetization for the larger $m = 2.75$~mg sample.  $M(\alpha)$ angle dependence of magnetization is $\pi$-periodic, see the inset to Fig.~\ref{fig5} (a), so the high-field linear branches are of different slopes in Fig.~\ref{fig5} (a) for M($\alpha$) minima and maxima, i.e. for  $\alpha$ = 35$^\circ$ and  $\alpha$ = 125$^\circ$, respectively. $M(H)$ magnetization is non-linear in low fields, so there is  low-field  $M(H)$ hysteresis with negative $M(H)$ slope within $\pm 2$~kOe,  as it is depicted in Fig.~\ref{fig5} (b). The behavior is qualitatively the same at 180~K temperature, see Fig.~\ref{fig5} (c), but the slopes of the linear high-field branches are diminished for this  $m = 2.75$~mg sample, in contrast to the smallest 1.39~mg  one in Fig.~\ref{fig3}. Thus, the temperature dependence of the $M(H)$ branches is due to the sample itself and it is not affected, e.g., by the experimental setup. For $M(\alpha)$ minima, magnetization is negative in low fields for both temperatures, see Fig.~\ref{fig5} (d).

\section{Discussion} \label{disc}

As a result, $M(\alpha)$ oscillates around zero value, so magnetization changes a sign for  $M(\alpha)$ maxima and minima, respectively. $M(H)$ is non-linear in low magnetic fields, but with almost linear high-field branches with positive ($M(\alpha)$ maxima) or negative ($M(\alpha)$ minima) slopes.  We also do not observe $\pi/3$ periodicity of magnetization for any sample orientation, despite it could be expected for the hexagonal CrSb structure.

For conventional antiferromagnets, one should expect slightly non-linear high-field branches with positive slope, and, possibly, with  small hysteresis. In particular,  high-field $M(H)$ branches are due to the antiferromagnetic spin-flop, which is simultaneous canting of two magnetic sublattices in external magnetic field much below the N\'eel vector reorientation field. In this case, small high-field $M(H)$ hysteresis reflects the antiferromagnetic domain configuration change. Even the low-field hysteresis is possible, e.g.,  for weak ferromagnetism~\cite{weak_ferro,weak_ferro1} or for the spin-orbit-induced spin polarization in altermagnets~\cite{alter_ferro,spin_ferro_soc,orlova_MnTe1,orlova_MnTe2}. 

However, for conventional antiferromagnets, the interplay between the positive and negative $M(H)$ slopes is impossible, as well as the negative magnetization values themselves.  Thus, $M(\alpha)$ oscillations around zero value definitely require several magnetic phases. 

We wish to emphasize, that the interplay between positive and negative $M(H)$ slopes can not originate from experimental setup. (i)  For our setup,  $M(\alpha)$ is not allowed to change a sign for homogeneous, single-phase sample magnetization: the magnetometer detector coils are fixed to the magnet pole caps (not to the sample holder), so, in finite external magnetic field, magnetization is always positive for ferromagnetic~\cite{cosns} and antiferromagnetic~\cite{orlova_MnTe1, orlova_MnTe2} samples, or negative for diamagnetic ones~\cite{gete}. (ii) Opposite $M(T)$ temperature dependences  for samples with different masses in Figs.~\ref{fig3} and~\ref{fig5} require to connect $M(T)$ with the samples themselves, but not with the experimental setup. (iii) The observed interplay is independently confirmed by the anomalous Hall effect measurements for the small single-crystal CrSb flake in Fig.~\ref{fig2} (c). 

CrSb belongs to a new class of altermagnetic materials with alternating spin splitting in the k-space~\cite{alter_common1, alter_common2, alter_mazin}. In this case, for the bulk magnetization, one could expect $M(\alpha)$ oscillations, so $M(\alpha)$ minima correspond to the directions of full spin compensation, while  $M(\alpha)$ maxima are due to the non-compensated  spins, as it has been demonstrated for MnTe altermagnet~\cite{orlova_MnTe1, orlova_MnTe2}. However, the slopes of high-field $M(H)$ branches are always positive in this case~\cite{orlova_MnTe1, orlova_MnTe2}. 

On the other hand, altermagnetic candidate CrSb reveals both altermagnetic and topological features with  Weyl surface states~\cite{Weyl alter2_CrSb,Weyl alter1_CrSb}. The current-carrying topological surface states lead to the diamagnetic response~\cite{orbital_mag2,orbital_mag_review,golub2025}, which dominates for the directions of full spin compensation in bulk altermagnetic spectrum. Thus, negative magnetization values for $M(\alpha)$ minima are due to the topological Weyl surface states in CrSb.

This conclusion is well supported by the experimental results. (i) The effect is mostly pronounced for the smaller samples with higher surface to  bulk ratio, see Figs.~\ref{fig2} and~\ref{fig5}. (ii) Since the diamagnetic response originates from the sample surface, it can be observed for any sample orientation, see  Figs.~\ref{fig2} and~\ref{fig3}  for the side and the the top mount position of the sample, respectively. (iii) Clear diamagnetic response is only possible for the full bulk spin compensation, which is affected by temperature, see  Figs.~\ref{fig4} and~\ref{fig5}. (iv) Due to the spin-momentum locking, topological surface states are spin-polarized, which is responsible for the  low-field hysteresis~\cite{gete}. Being induced by the topological surface states~\cite{gete}, the hysteresis is always characterized by negative (diamagnetic-like) $M(H)$ slope,

Thus, we demonstrate interplay between  altermagnetic  bulk and topological surface magnetizations for the  altermagnetic candidate CrSb.

\section{Conclusion}

As a conclusion, we experimentally investigate the angle dependence of magnetization   $M(\alpha)$ for single crystals of CrSb. CrSb belongs to a new class of altermagnetic materials, the small net magnetization is accompanied by alternating spin splitting in the k-space. In addition,  CrSb reveals also topological features with  Weyl surface states originating from bulk band topology. We observe, that $M(\alpha)$ oscillates around zero value, so magnetization is positive for $M(\alpha)$ maxima and it is negative for $M(\alpha)$ minima. The magnetization reversal curves $M(H)$ are non-linear with low-field hysteresis, but with almost linear high-field branches. The slope of the linear branches well correlates with $M(\alpha)$ oscillations, so it is positive for $M(\alpha)$ maxima and negative for $M(\alpha)$ minima. We demonstrate, that the interplay between the positive and the negative $M(H)$ slopes originates from several magnetic phases in CrSb. In particular, current-carrying topological surface states are responsible for the diamagnetic-like $M(H)$ negative slope, which dominates for the directions of full spin compensation in the bulk CrSb altermagnetic spectrum. Due to the spin-momentum locking, topological surface states are spin-polarized, which is responsible for the  low-field hysteresis.  Thus, we experimentally  demonstrate both the altermagnetic  bulk and the topological surface magnetizations for the  altermagnetic candidate CrSb.

\acknowledgments

We wish to thank S.S~Khasanov for X-ray sample characterization and V.A. Zyuzin for valuable discussions.

\end{document}